\documentclass[11pt]{article}

\usepackage[a4paper,margin=1in]{geometry}
\usepackage{graphicx}      
\usepackage{natbib}        
\usepackage{amsmath}       
\usepackage{amssymb} 
\usepackage{subcaption}  
\usepackage{booktabs} 

\usepackage[english]{babel}
\usepackage{tabularx}
\usepackage{algorithm}
\usepackage{algpseudocode}
\usepackage[font=small,labelfont=bf]{caption}
\usepackage{authblk}
\usepackage{hyperref}

\begin{document}

\title{Bayesian Optimization of a Multi-Product Chemical Reactor Using Composite Models and Partial Physics Knowledge} 

\author[1]{Liqiu Dong} 
\author[2]{Marta Zag\'orowska} 
\author[1]{Mehmet Mercang\"oz}

\affil[1]{Department of Chemical Engineering, Imperial College London,
London, UK (e-mail: liqiu.dong22@imperial.ac.uk).}
\affil[2]{DCSC, Delft University of Technology, Delft, Netherlands}

\date{}

\maketitle

\begin{center}
\small
\textcopyright\ 2026 the authors.
This work has been accepted to IFAC for publication under a Creative Commons Licence CC-BY-NC-ND.
\end{center}

\begin{abstract}
We study data-driven real-time economic optimization of a multi-product chemical reactor when no reliable first-principles model is available beyond a steady-state energy balance. Instead of learning the economic objective directly as a black-box function, we use a composite formulation in which Gaussian process (GP) models predict physically meaningful outputs, including product concentrations and reactor temperature, while profit is computed analytically from these predictions together with raw-material, product, and utility prices. This preserves the structure of the economic objective, makes it parametric in changing prices without needing retraining, and allows candidate operating points to be checked against the available energy balance through a physics residual. The GPs also provide predictive uncertainty, which is exploited in a Bayesian optimization (BO) framework both for data-efficient exploration and for conservative enforcement of the reactor temperature constraint through an upper confidence bound. The acquisition function additionally penalizes large energy-balance mismatch obtained by substituting the GP-predicted outputs and candidate inputs into the available steady-state energy balance. The approach is demonstrated on a benchmark simulation of a non-isothermal multi-product reactor. Relative to a trust-region safe BO implementation, the proposed method achieves better simulated economic performance within the available iteration budget. Relative to a purely data-driven BO approach that does not use the available physics information, it avoids reactor temperature constraint violations.
\end{abstract}



\section{Introduction}

Real-time optimization (RTO) is used to choose operating points for process systems to improve economic performance while respecting operational and safety limits. In practice, however, the quality of the resulting operating points depends strongly on the available process model, and developing reliable first-principles models for complex systems can be costly and difficult. This motivates approaches that combine partial physical knowledge, such as conservation laws, with data-driven optimization. In this work, we pursue such a strategy for the economic optimization of a multi-product chemical reactor.

Physics-informed machine learning has emerged as a powerful paradigm for embedding known physical laws and governing equations into data-driven models, often improving reliability and data efficiency relative to purely empirical approaches \citep{karniadakisPhysicsinformedMachineLearning2021}. In parallel, Bayesian optimization (BO) has become a widely used methodology for optimizing expensive black-box systems under limited evaluation budgets. Gaussian process (GP) surrogates are particularly attractive in this setting because they provide both predictions and uncertainty estimates, enabling data-efficient sequential exploration \citep{10.7551/mitpress/3206.001.0001, krishnamoorthyModelfreeRealtimeOptimization2023}. 

A key challenge arises when BO is applied to safety-critical systems. Purely data-driven exploration may drive the process into unsafe operating regions before the optimizer has learned where the constraints lie. For this reason, safe BO methods incorporate uncertainty-aware mechanisms into the acquisition function so that exploration remains conservative near operational limits \citep{suiSafeExplorationOptimization2015, dong2024real, Korkmaz_2023}. A promising extension is to further enrich BO with available physical knowledge. Recent physics-informed BO studies have shown that known governing relations can guide the search toward physically plausible regions and improve efficiency, although most existing applications have focused on offline design problems rather than online chemical process optimization \citep{khatamsazPhysicsInformedBayesian2023, kobayashiPhysicsinformedBayesianOptimization2025}. Rather than learning the economic objective directly as a black-box function of the manipulated variables, we adopt a \emph{composite} formulation: multiple GP models map the inputs to physically meaningful steady-state outputs and plant profit is then computed analytically from these outputs together with exogenous economic parameters. This preserves the structure of the economic objective and keeps the dependence on economic conditions explicit. To enforce safety, we penalize candidate inputs whose upper confidence bound violates safety constraints. In addition, we exploit partial first-principles knowledge by substituting the GP predictions into the available steady-state energy balance to define a \emph{physics residual}. This residual is incorporated into the acquisition function to discourage operating points that are inconsistent with the known physics while preserving BO's global exploration capability.

The main contributions of this paper are as follows: (i) a \emph{composite} BO formulation in which profit is computed analytically from GP models of the steady-state outputs, allowing economic parameters to change without retraining; (ii) a physics-residual penalty based on the steady-state energy balance, embedded in the acquisition function to discourage physically inconsistent sampling; and (iii) a demonstration on a non-isothermal multi-product reactor, showing improved economic performance over a trust-region safe BO baseline and avoidance of temperature-constraint violations relative to a purely data-driven BO baseline.

In the remainder of this paper, Section 2 describes the preliminaries for safe data-driven RTO under uncertainty. Section 3 introduces the proposed methodology and the acquisition function that integrates physics residuals as a penalty term. Section 4 describes the case study and the RTO problem formulation, while Section 5 presents and discussed the results compared to other learning-based approaches. Finally, Section 6 provides a conclusion.

\section{Background}

\subsection{Bayesian optimization}
Bayesian optimization (BO) in its basic form addresses unconstrained optimization problems in which function evaluations are expensive and the analytical form of the objective is unavailable:
\begin{equation}
    \mathbf{u}^* \in \arg \min_{\mathbf{u} \in \mathcal{U}} \mathrm{J}(\mathbf{u}),
    \label{eq:bo_problem}
\end{equation}
where $\mathbf{u} \in \mathcal{U} \subset \mathbb{R}^{n_u}$ is the decision vector and $\mathrm{J}(\mathbf{u})$ is an unknown black-box objective. BO iteratively seeks near-optimal solutions by combining a probabilistic surrogate model of $\textrm{J}$ with an acquisition function $\alpha_n(\mathbf{u})$ \citep{10.7551/mitpress/3206.001.0001}:
\begin{equation}
    \mathbf{u}_{n+1} \in \arg \min_{\mathbf{u} \in \mathcal{U}} \alpha_n(\mathbf{u}),
    \label{eq:af_optimization}
\end{equation}
The surrogate models are usually obtained from measured data, for example in the form of Gaussian processes \citep{Srinivas2012,10.7551/mitpress/3206.001.0001}.

\subsection{Safe BO in process control} 
\label{sec:SafeBO}
\subsubsection{Composite BO} In process optimization, the quantities of interest are typically functions of the steady-state plant outputs $\mathbf{x}:\mathbb{R}^{n_u} \rightarrow \mathbb{R}^{n_x},$ rather than direct observations of a scalar objective. The economic performance can therefore be written as a known function of the outputs and inputs, while operational limits are imposed on selected output components. This leads to a \emph{composite} BO formulation in which surrogate models are built for the steady-state outputs, while the objective is evaluated analytically from those predictions. 
In this work, a separate GP is used for each steady-state output component $x_i(\mathbf{u})$, $i=1,\dots,n_x$. After $n$ iterations, the available data are:
\begin{equation}
    \mathcal{D}_n = \{(\mathbf{u}_k,\mathbf{x}_k)\}_{k=1}^n,
\end{equation}
For output $i$, the latent input--output relation is modeled as:
\begin{equation}
    x_i(\mathbf{u}) \sim \mathcal{GP}(0,k_i(\mathbf{u},\mathbf{u}')),
\end{equation}
with noisy observations:
\begin{equation}
    x_{i,k}^{\mathrm{obs}} = x_i(\mathbf{u}_k) + \varepsilon_{i,k},
    \qquad
    \varepsilon_{i,k} \sim \mathcal{N}(0,\sigma_{n,i}^2).
\end{equation}
For any candidate input $\mathbf{u}$, the GP posterior is Gaussian:
\begin{equation}
    x_i(\mathbf{u}) \mid \mathcal{D}_n \sim \mathcal{N}\!\left(\mu_i(\mathbf{u}),\sigma_i^2(\mathbf{u})\right),
\end{equation}
with mean and variance:
\begin{align}
    \mu_i(\mathbf{u})
    &= \mathbf{k}_i(\mathbf{u})^\top
    \left(\mathbf{K}_i + \sigma_{n,i}^2 \mathbf{I}\right)^{-1}
    \mathbf{x}^{(n)}_i, \\
    \sigma_i^2(\mathbf{u})
    &= k_i(\mathbf{u},\mathbf{u})
    - \mathbf{k}_i(\mathbf{u})^\top
    \left(\mathbf{K}_i + \sigma_{n,i}^2 \mathbf{I}\right)^{-1}
    \mathbf{k}_i(\mathbf{u}),
\end{align}
where $\mathbf{U}_n=[\mathbf{u}_1,\dots,\mathbf{u}_n]^\top$, $\mathbf{x}^{(n)}_i=[x_{i,1}^{\mathrm{obs}},\dots,x_{i,n}^{\mathrm{obs}}]^\top$, $\mathbf{K}_i = K_i(\mathbf{U}_n,\mathbf{U}_n)$, and $\mathbf{k}_i(\mathbf{u}) = K_i(\mathbf{U}_n,\mathbf{u})$. Collecting the GP mean predictions gives:
\begin{equation}
    \boldsymbol{\mu}(\mathbf{u})
    :=
    [\mu_1(\mathbf{u}),\dots,\mu_{n_x}(\mathbf{u})]^\top.
    \label{eq: mean_definition}
\end{equation}

\subsubsection{Constrained BO} Operational limits are captured by safety constraints, often written as:
\begin{equation}
\label{eq:SefetyConstraint}
   c(\mathbf{x}, \mathbf{u})\leq 0
\end{equation}
where $c()$ is a function of outputs $\mathbf{x}$ and inputs $\mathbf{u}$, for instance $c(\mathbf{x},\mathbf{u})=\mathbf{x}-\mathbf{x}^{\max}$ where $\mathbf{x}^{\max}$ is constant.

When constraints \eqref{eq:SefetyConstraint} are learned from limited data, a common BO strategy is to use uncertainty-aware conservative estimates \citep{suiSafeExplorationOptimization2015, dong2024real, Korkmaz_2023}. For an upper-bound constraint, this leads to the upper confidence bound:
\begin{equation}
    \hat{c}^{\mathrm{UCB}}(\mathbf{u})
    :=
    \mu_c(\mathbf{u}) + \beta \sigma_c(\mathbf{u}),
    \label{eq:ucb_temperature}
\end{equation}
where $\mu_c(\mathbf{u})$ and $\sigma_c(\mathbf{u})$ are the GP posterior mean and standard deviation for the constraint, and $\beta>0$ controls conservatism \citep{Srinivas2012}. Inputs for which $\hat{c}^{\mathrm{UCB}}(\mathbf{u})$ violates \eqref{eq:SefetyConstraint} are treated as risky and can be penalized or excluded within the acquisition function.

\subsubsection{Partial physics information}In addition to learned constraints, partial first-principles knowledge may also be available. In many chemical-process applications, a complete model of the plant is unavailable, but specific physical relationships, such as steady-state mass or energy balances, are still known. These relationships can be used to check whether surrogate predictions are physically consistent \citep{khatamsazPhysicsInformedBayesian2023, kobayashiPhysicsinformedBayesianOptimization2025, dongPICoFBilevelOptimization2024}. If a known steady-state relation is:
\begin{equation}
    \mathcal{F}_p(\mathbf{x},\mathbf{u}) = 0,
    \label{eq:physics_governing_method}
\end{equation}
then a model--physics residual can be defined by substituting the GP mean predictions \eqref{eq: mean_definition} into \eqref{eq:physics_governing_method}. A large residual indicates that the candidate operating point is inconsistent with the available physics. The residual is evaluated at the GP mean predictions rather than through uncertainty bounds, since unlike the safety constraint it is not a one-sided risk quantity but a measure of inconsistency between the surrogate prediction and the available physical knowledge.

These two elements, conservative handling of uncertain safety constraints and residual-based use of partial physics knowledge, provide the basis for the methodology developed in the next section.

\section{Methodology}

We consider a system with manipulated inputs $\mathbf{u} \in \mathcal{U} \subset \mathbb{R}^{n_u}$. For any candidate input $\mathbf{u}$, the plant reaches a steady state characterized by the output vector $\mathbf{x}$.
Rather than learning the economic objective directly as a scalar black-box function of $\mathbf{u}$, we exploit the fact that the objective can be written analytically as a known function of the steady-state outputs and inputs $\mathrm{J}(\mathbf{x},\mathbf{u};\boldsymbol{\theta})$, where $\boldsymbol{\theta}$ denotes exogenous economic parameters such as raw-material prices, product prices, and utility costs. The system is also subject to safety constraints of the form \eqref{eq:SefetyConstraint}
and must satisfy physical relations \eqref{eq:physics_governing_method}. To model the steady-state plant response, we employ a set of independent GPs, one for each output component $x_i(\mathbf{u})$, $i=1,\dots,n_x$, as indicated in Section \ref{sec:SafeBO}. For any candidate input $\mathbf{u}$, the GP models provide posterior means and standard deviations for each output component defined by \eqref{eq: mean_definition}. 

The economic objective is then evaluated in composite form at the GP mean predictions, i.e.\ through $\mathrm{J}(\boldsymbol{\mu}(\mathbf{u}),\mathbf{u};\boldsymbol{\theta})$. This preserves the analytical structure of the objective and keeps the dependence on the economic parameters explicit.

\subsection{Composite acquisition function}
We propose a composite acquisition function, $\alpha_{PI}(\mathbf{u})$, that balances economic exploitation, statistical exploration, conservative safety handling, and consistency with the available physics:
\begin{equation}
\begin{aligned}
\alpha_{PI}(\mathbf{u}) =\;&
\underbrace{\mathrm{J}(\boldsymbol{\mu}(\mathbf{u}), \mathbf{u};\boldsymbol{\theta})}_{\text{Exploitation}}
\;+\;
\underbrace{\left(- z \sum_{i=1}^{n_x} \sigma_i(\mathbf{u})\right)}_{\text{Exploration}} 
+\;
\underbrace{\lambda\, P_{\mathrm{safe}}(\mathbf{u})}_{\text{Safety}}
\;+\;
\underbrace{\gamma\, P_{\mathrm{phys}}(\mathbf{u})}_{\text{Physics}} .
\end{aligned}
\label{eq:pi_acquisition}
\end{equation}
where $z \ge 0$, $\lambda \ge 0$, and $\gamma \ge 0$ are tuning parameters. The first term, Exploitation, favors economically attractive operating points, the second, Exploration, promotes exploration in regions of high predictive uncertainty, the third term, $P_{\mathrm{safe}}$, penalizes candidate points that are likely to violate safety constraints, and the fourth term,$P_{\mathrm{phys}}$, penalizes inconsistency with the available physical relations. The selection of acquisition-function weights is problem-dependent. Even in standard GP-UCB, the exploration parameter controls the exploitation--exploration trade-off and is often chosen heuristically. Automatic tuning of acquisition functions has been studied, but no universal rule exists for composite acquisitions that combine economic and safety \citep{Srinivas2012}. 

\subsubsection{Uncertainty-aware safety constraint}
Using \eqref{eq:ucb_temperature}, the penalty term is written as:
\begin{equation}
    P_{\mathrm{safe}}(\mathbf{u}) =
    \sum_{j=1}^{n_c} \frac{1}{2}
    \left[
    \tanh\!\left(
    \alpha_{\mathrm{safe}} \hat{c}_j^{\mathrm{UCB}}(\mathbf{u})
    \right) + 1
    \right],
    \label{eq:safe_penalty_method}
\end{equation}
where $\alpha_{\mathrm{safe}} > 0$ controls the sharpness of the transition around the constraint boundary, $n_c$ is the number of safety constraints and $\hat{c}_j^{\mathrm{UCB}}(\mathbf{u})$ is the conservative upper-confidence estimate of the $j$-th constraint. 

\subsubsection{Physics-informed residual penalty}
To incorporate partial first-principles knowledge, we substitute the GP mean predictions into the available steady-state physical relation \eqref{eq:physics_governing_method}. This gives the model--physics residual:
\begin{equation}
    \mathcal{R}_p(\mathbf{u})
 = \mathcal{F}_p(\boldsymbol{\mu}(\mathbf{u}), \mathbf{u}),
\end{equation}
from which the physics penalty is defined as:
\begin{equation}
    P_{\mathrm{phys}}(\mathbf{u}) = \| \mathcal{R}_p(\mathbf{u})
 \|.
    \label{eq:physics_penalty_method}
\end{equation}

\subsection{Sequential optimization procedure}
The proposed optimization loop is summarized in Algorithm~\ref{alg:pibo}. Starting from an initial dataset of safe operating points $\mathcal{D}_0$, the independent GP models are fitted to the measured steady-state outputs. At each BO iteration, the next candidate input is obtained from \eqref{eq:af_optimization} with \eqref{eq:pi_acquisition}. The plant is evaluated at $\mathbf{u}_{n+1}$, the resulting steady-state output vector is added to the dataset $\mathcal{D}$, and the GP models are updated. This process is repeated until the available evaluation budget is exhausted. 

\begin{algorithm}[t]
\caption{Physics-residual-guided BO}
\label{alg:pibo}
\begin{algorithmic}[1]
\Require a safe dataset $\mathcal{D}_0$, an input domain $\mathcal{U}$, an evaluation budget $N$, 
fixed acquisition-function weights
\For{$n = n_0,\ldots,N-1$}
    \State Fit GPs for $x_i(\mathbf{u})$, $i=1,\ldots,n_x$, using $\mathcal{D}_n$
    \State Compute $\boldsymbol{\mu}(\mathbf{u})$ and $\sigma_i(\mathbf{u})$ from the GP posteriors
    \State Compute uncertainty-aware safety penalty $P_{\mathrm{safe}}(\mathbf{u})$
    \State Compute residual $\mathcal{R}_p(\mathbf{u})$ and penalty $P_{\mathrm{phys}}(\mathbf{u})$
    \State Form physics-informed acquisition function $\alpha_{PI}(\mathbf{u})$
    \State Select the next operating point using \eqref{eq:af_optimization}
    \State Apply $\mathbf{u}_{n+1}$ to obtain $\mathbf{x}_{n+1}$
    \State Update
    $
        \mathcal{D}_{n+1}
        =
        \mathcal{D}_n \cup \{(\mathbf{u}_{n+1},\mathbf{x}_{n+1})\}
    $
\EndFor
\State \Return best feasible evaluated point according to $\mathrm{J}(\mathbf{x},\mathbf{u};\boldsymbol{\theta})$
\end{algorithmic}
\end{algorithm}

\section{Case Study}

We evaluate the proposed method on a non-isothermal continuous stirred-tank reactor (CSTR), following the standard structure of CSTR benchmark models with Arrhenius kinetics, competing reaction pathways, and heat-transfer dynamics \citep{article_CSTR}. This system exhibits strong nonlinearity, competing parallel reactions, and a critical temperature safety limit, which together create a challenging non-convex optimization problem.

\subsection{Reactor system and reaction kinetics}
\begin{figure}[h!]
  \centering
  \includegraphics[width=0.5\columnwidth]{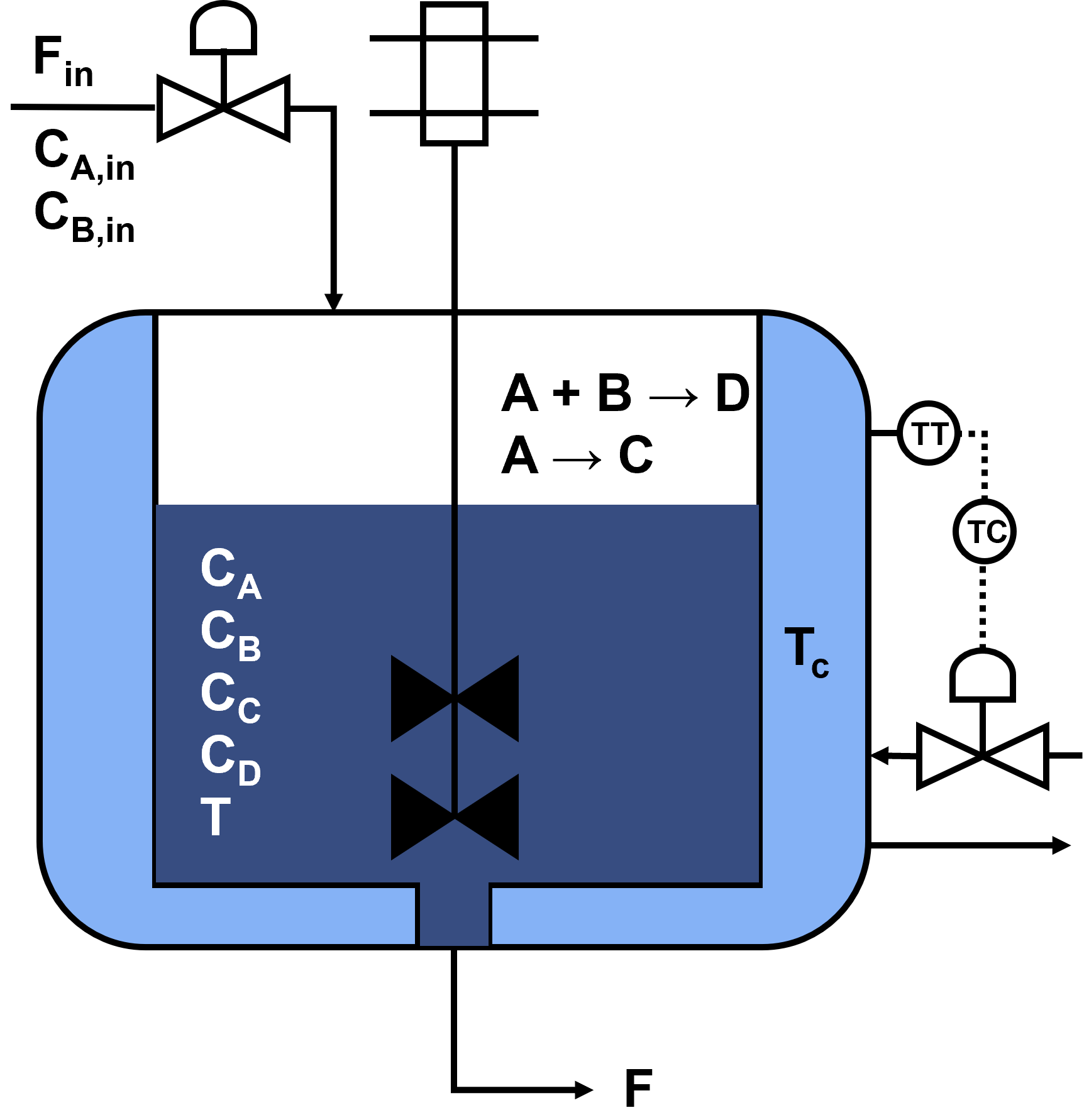}
  \caption{A schematic of the Continuous Stirred-Tank Reactor (CSTR).}
  \label{fig:my_cstr}
\end{figure}

As shown in Fig.~\ref{fig:my_cstr}, the reactor involves two irreversible and exothermic parallel reactions:
\begin{equation}
    A + B \xrightarrow{r_1} D,
    \qquad
    A \xrightarrow{r_2} C.
\end{equation}
Here, $D$ is the desired product, while $C$ is a lower-value side product. The reaction rates are assumed to follow elementary kinetics:
\begin{equation}
    r_1 = k_1 C_A C_B,
    \qquad
    r_2 = k_2 C_A,
    \label{eq:reaction_rates}
\end{equation}
with Arrhenius rate constants:
\begin{equation}
    k_j(T) = k_{0,j} \exp\left(\frac{-E_{A,j}}{RT}\right),
    \qquad j=1,2,
    \label{eq:arrhenius}
\end{equation}
where $k_{0,j}$ is the pre-exponential factor, $E_{A,j}$ is the activation energy, $T$ is the reactor temperature, and $R$ is the universal gas constant.

\subsection{Dynamic simulation model}
The plant is simulated using five coupled ordinary differential equations representing the species and energy balances. Assuming constant volume $V$, density $\rho$, and heat capacity $c_p$, the model is:
\begin{align}
    \frac{dC_A}{dt} &= \frac{F}{V}(C_{A,\mathrm{in}} - C_A) - r_1 - r_2,
    \label{eq:ode_Ca}\\
    \frac{dC_B}{dt} &= \frac{F}{V}(C_{B,\mathrm{in}} - C_B) - r_1,
    \label{eq:ode_Cb}\\
    \frac{dC_C}{dt} &= -\frac{F}{V} C_C + r_2,
    \label{eq:ode_Cc}\\
    \frac{dC_D}{dt} &= -\frac{F}{V} C_D + r_1,
    \label{eq:ode_Cd}\\
    \frac{dT}{dt} &= \frac{F}{V}(T_{\mathrm{in}} - T)
    + \frac{1}{\rho c_p} \sum_{j=1}^{2} (-\Delta H_j) r_j
    + \frac{UA}{V \rho c_p}(T_c - T).
    \label{eq:ode_T}
\end{align}
where $C_i$ denotes the concentration of species $i$, and $T_c$ is the coolant temperature. The inlet concentrations are parameterized by the total inlet concentration $C_{\mathrm{tot}}$ and the feed fraction of species $A$, denoted by $f_A$:
\begin{equation}
    C_{A,\mathrm{in}} = f_A C_{\mathrm{tot}},
    \qquad
    C_{B,\mathrm{in}} = (1-f_A) C_{\mathrm{tot}}.
\end{equation}
All remaining model parameters are listed in Table~\ref{tab:model_parameters}. The model \eqref{eq:ode_Ca}-\eqref{eq:ode_T} will be used to simulate the real plant and obtain steady-state measurements used for the surrogate models. For optimization purposes, we assume that only the energy balance equation \eqref{eq:ode_T} is available.

\subsection{Steady-state optimization problem}
The decision vector is:
\begin{equation}
    \mathbf{u} = [f_A,\; T_c^{\mathrm{scaled}}]^\top \in [0,1]^2,
\end{equation}
where $f_A$ is the inlet fraction of species $A$ and $T_c^{\mathrm{scaled}}$ is a normalized coolant-temperature variable. The physical coolant temperature is recovered by linear scaling:
\begin{equation}
    T_c = T_{c,\min} + T_c^{\mathrm{scaled}} (T_{c,\max} - T_{c,\min}),
\end{equation}
with $T_{c,\min}=273$ K and $T_{c,\max}=623$ K.

For any input $\mathbf{u}$, the plant evaluation returns the steady-state output vector:
\begin{equation}
    \mathbf{x}(\mathbf{u}) = [C_A,\; C_B,\; C_C,\; C_D,\; T]^\top.
\end{equation}
The optimization problem is therefore formulated as minimization of the objective:
\begin{equation}
    \min_{\mathbf{u} \in [0,1]^2} \mathrm{J}(\mathbf{x}(\mathbf{u}),\mathbf{u}),
\end{equation}
where the profit function is (minus sign indicates profit maximization):
\begin{equation}
\begin{aligned}
    \mathrm{J}(\mathbf{x},\mathbf{u}) =\;&
    -(F \left(P_C C_C + P_D C_D - P_A C_{A,\mathrm{in}} - P_B C_{B,\mathrm{in}}\right) 
    - P_{\mathrm{Heat}}\, UA\, \left|T_c(\mathbf{u}) - T\right|).
\end{aligned}
\label{eq:profit}
\end{equation}
The economic parameters used in the profit calculation are $P_A = 5.0$~\$/mol, $P_B = 2.0$~\$/mol, $P_C = 7.0$~\$/mol, $P_D = 19.0$~\$/mol, and $P_{\mathrm{Heat}} = 1\times10^{-5}$~\$/(kW$\cdot$ s).

The reactor is subject to the temperature safety constraint:
\begin{equation}
    T \le T_{\max},
    \qquad
    T_{\max} = 670.0~\mathrm{K}.
    \label{eq:constraint}
\end{equation}
\begin{figure*}[h] 
    \centering
    \begin{subfigure}[b]{0.32\textwidth}
        \centering
        \includegraphics[width=\linewidth]{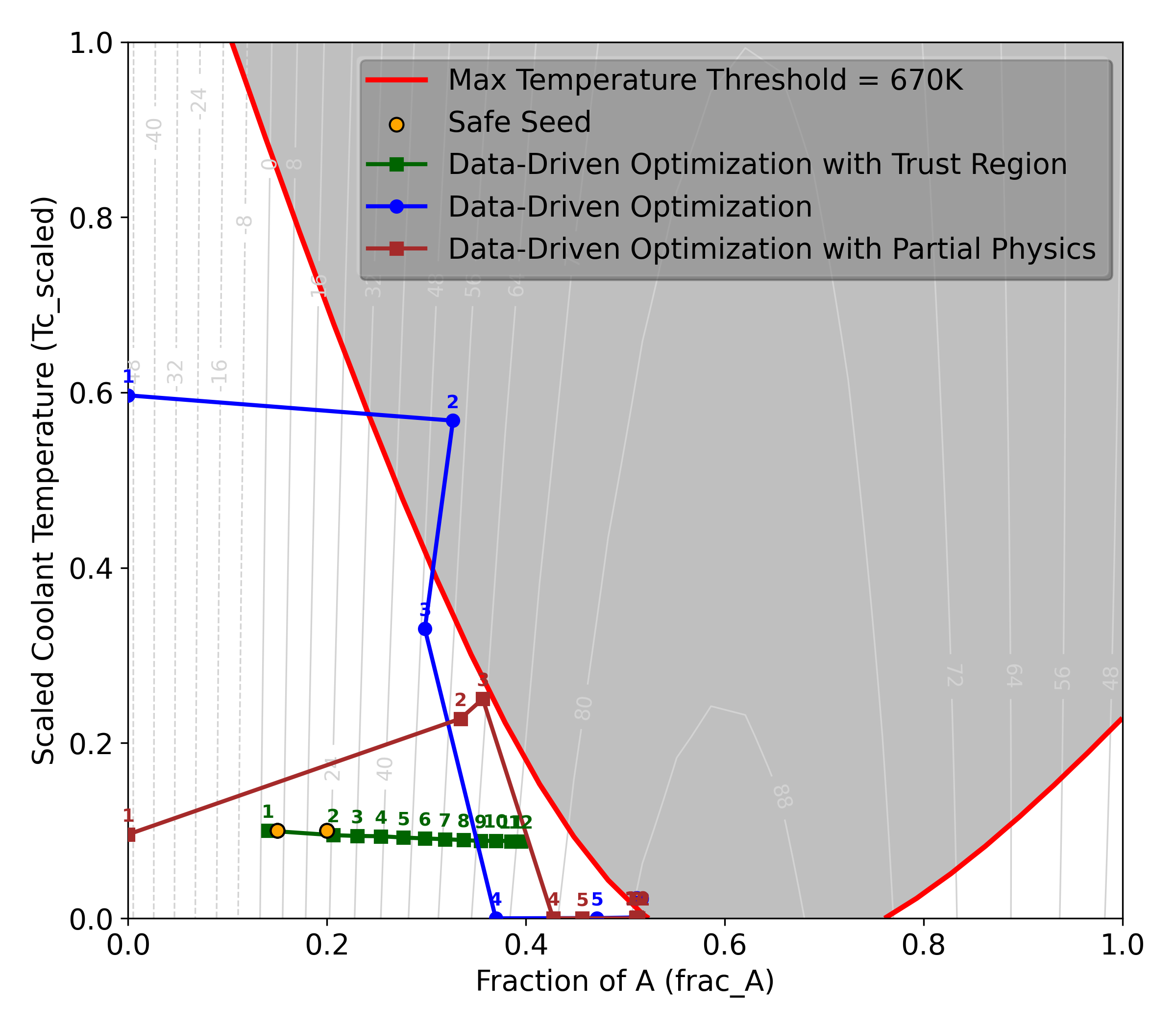} 
        \caption{Seed Scenario I}
        \label{fig:traj_a}
    \end{subfigure}
    \hfill
    \begin{subfigure}[b]{0.32\textwidth}
        \centering
        \includegraphics[width=\linewidth]{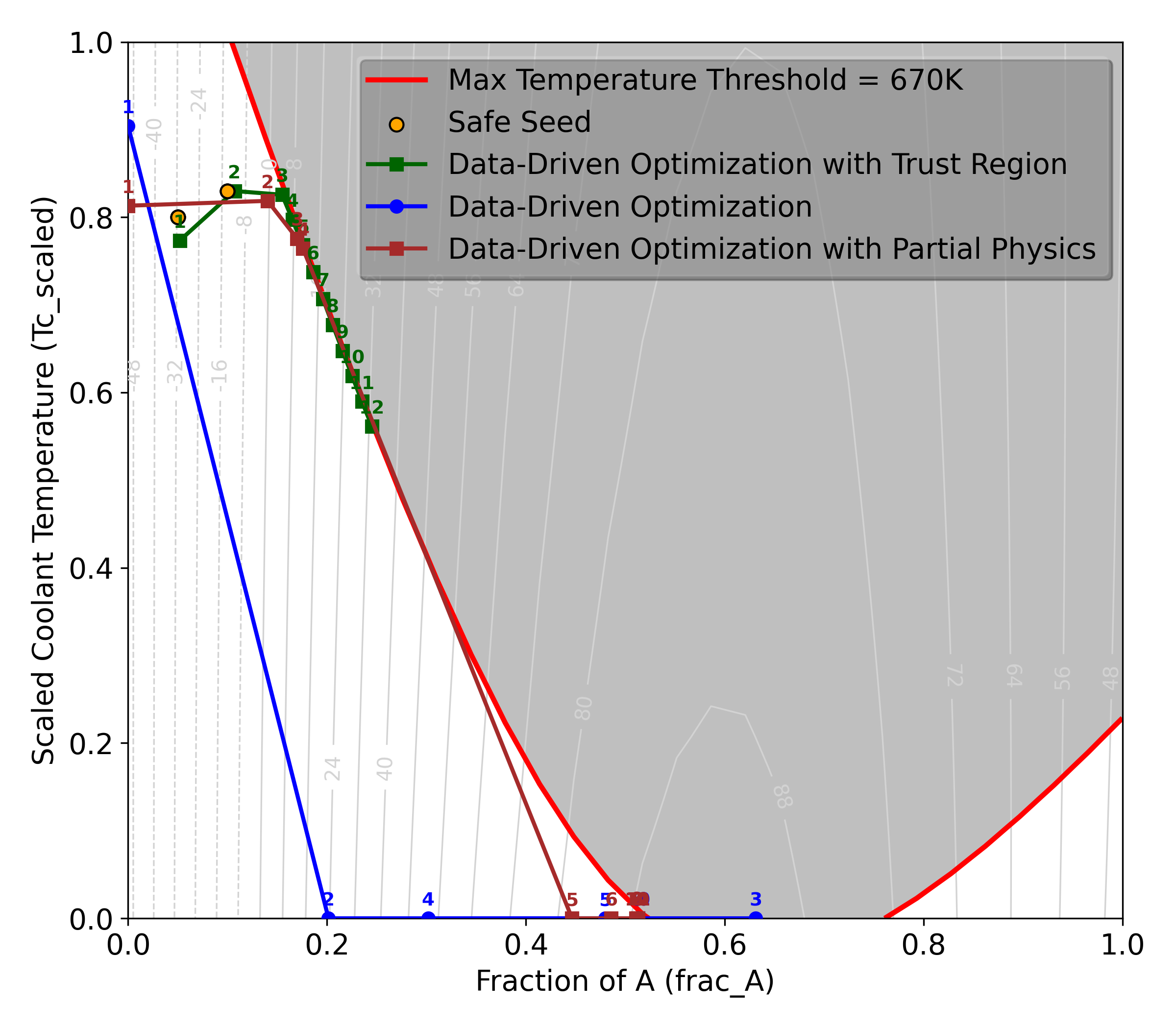} 
        \caption{Seed Scenario II }
        \label{fig:traj_b}
    \end{subfigure}
    \hfill
    \begin{subfigure}[b]{0.32\textwidth}
        \centering
        \includegraphics[width=\linewidth]{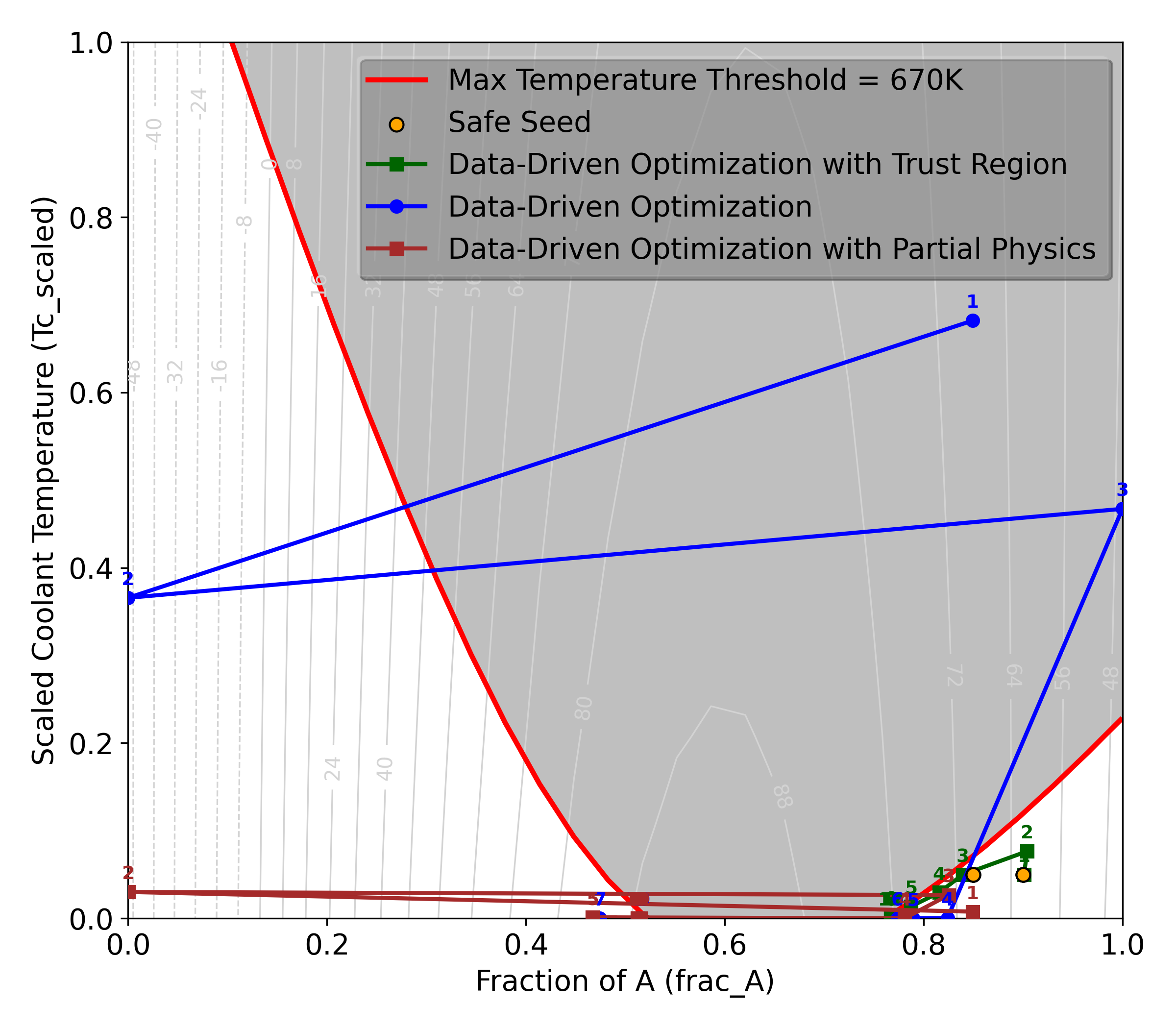} 
        \caption{Seed Scenario III}
        \label{fig:traj_c}
    \end{subfigure}
    \caption{Optimization trajectories overlaid on the ground-truth economic profit landscape. The dark gray region denotes the infeasible operating area ($T > 670$ K). Blue trajectories represent the Baseline algorithm. Green trajectories represent the trust-region method. Red trajectories represent the proposed algorithm.}
    \label{fig:trajectories}
\end{figure*}

\subsection{Case-specific acquisition function}
The steady-state outputs $\mathbf{x}(\mathbf{u}) = [C_A,\; C_B,\; C_C,\; C_D,\; T]^\top$ are modeled using five independent GPs. For a candidate input $\mathbf{u} = [f_A,\; T_c^{\mathrm{scaled}}]^\top$, the conservative temperature estimate is defined as:
\begin{equation}
    \hat{T}^{\mathrm{UCB}}(\mathbf{u}) = \mu_T(\mathbf{u}) + 2\sigma_T(\mathbf{u}),
    \label{eq:Tucb}
\end{equation}
and the CSTR-specific acquisition function is written as:
\begin{equation}
\begin{aligned}
\alpha_{\mathrm{CSTR}}(\mathbf{u}) =\;&
\mathrm{J}\!\left(\boldsymbol{\mu}(\mathbf{u}),\mathbf{u}\right)
- z \sum_{i=1}^{5} \sigma_i(\mathbf{u}) 
+ \frac{\lambda}{2}
\left[
\tanh\!\left(
\alpha_{\mathrm{safe}}
\bigl(\hat{T}^{\mathrm{UCB}}(\mathbf{u}) - T_{\max}\bigr)
\right) + 1
\right]
+ \gamma\, \mathcal{R}_p(\mathbf{u}),
\end{aligned}
\label{eq:acq_cstr_exact}
\end{equation}
where $\mathrm{J}(\boldsymbol{\mu}(\mathbf{u}),\mathbf{u})$ denotes the profit evaluated from the GP mean predictions using \eqref{eq:profit}. The second term promotes exploration, the third penalizes candidate points that may violate the temperature constraint, and the final term penalizes inconsistency with the available steady-state energy balance through the energy balance residual $\mathcal{R}_p$, computed as:
\begin{equation}
\begin{aligned}
\mathcal{R}_p(\mathbf{u}) = \Bigg|
&\frac{F}{V}\bigl(T_{\mathrm{in}}-\mu_T(\mathbf{u})\bigr)
+ \frac{1}{\rho c_p}
\sum_{j=1}^{2} (-\Delta H_j)\,\hat{r}_j(\mathbf{u})
+ \frac{UA}{V\rho c_p}
\bigl(T_c(\mathbf{u})-\mu_T(\mathbf{u})\bigr)
\Bigg|,
\end{aligned}
\label{eq:energy_residual_cstr}
\end{equation}
where the reaction rates implied by the GP mean predictions are:
\begin{equation}
\begin{aligned}
    \hat{r}_1(\mathbf{u}) = &{}
    k_1\!\bigl(\mu_T(\mathbf{u})\bigr)\,
    \mu_{C_A}(\mathbf{u})\,
    \mu_{C_B}(\mathbf{u}),\\
    \hat{r}_2(\mathbf{u}) = &{}
    k_2\!\bigl(\mu_T(\mathbf{u})\bigr)\,
    \mu_{C_A}(\mathbf{u}).
\end{aligned}
\end{equation}
Thus, $\mathcal{R}_p(\mathbf{u})$ is obtained by setting $dT/dt=0$ in \eqref{eq:ode_T} and substituting the GP mean predictions into the steady-state energy balance.

For the results reported in this paper, the acquisition-function hyperparameters were set to
$z = 2.5$, $\lambda = 10^9$, $\gamma = 2$
and
$\alpha_{\mathrm{safe}} = 5$.
GP models were implemented with scikit-learn's \texttt{GaussianProcessRegressor}, using an RBF kernel and standardized inputs and outputs.

\begin{table}[htbp]
\centering
\caption{Parameter values used in the CSTR case study}
\label{tab:model_parameters}
\begin{tabularx}{\columnwidth}{@{} X c c l @{}} 
\toprule
\textbf{Parameter Description} & \textbf{Symbol} & \textbf{Value} & \textbf{Unit} \\
\midrule
\multicolumn{4}{@{}l}{\textit{Reaction Kinetics}} \\
Pre-exponential factor (Rxn 1) & $k_{0,1}$ & 1 & L/(mol$\cdot$s) \\
Pre-exponential factor (Rxn 2) & $k_{0,2}$ & 10 & 1/s \\
Activation energy (Rxn 1) & $E_{A,1}$ & 21500 & J/mol \\
Activation energy (Rxn 2) & $E_{A,2}$ & 91500 & J/mol \\
Universal gas constant & $R$ & 8.314 & J/(mol$\cdot$K) \\
\midrule
\multicolumn{4}{@{}l}{\textit{Thermodynamics}} \\
Enthalpy of reaction 1 & $\Delta H_1$ & -1300 & kJ/mol \\
Enthalpy of reaction 2 & $\Delta H_2$ & -400 & kJ/mol \\
Specific heat capacity & $c_p$ & 1 & kJ/(kg$\cdot$K) \\
Liquid density & $\rho$ & 1 & kg/L \\
\midrule
\multicolumn{4}{@{}l}{\textit{Process Operations}} \\
Reactor volume & $V$ & 1000 & L \\
Volumetric flow rate & $F$ & 5 & L/s \\
Heat-transfer coefficient & $UA$ & 25 & kW/K \\
Total inlet concentration & $C_{\mathrm{tot}}$ & 2.0 & mol/L \\
Inlet temperature & $T_{\mathrm{in}}$ & 300 & K \\
\bottomrule
\end{tabularx}
\end{table}

\section{Results and Discussion}

To visualize the optimization landscape, a high-fidelity reference map was generated by simulating the reactor on a fine $30\times 30$ grid over the input space. Figure~\ref{fig:trajectories} shows the resulting profit landscape, with the infeasible region defined by $T > T_{\max}$ highlighted in dark gray. The figure also overlays optimization trajectories from three distinct safe initializations. The baseline method (blue trajectories) is guided only by the GP mean and uncertainty. As a result, it explores near or within the high-temperature region in order to learn the constraint boundary, leading to several evaluations in an unsafe and physically inconsistent area. By contrast, the proposed method (red trajectories) uses the steady-state energy-balance residual to identify such regions as implausible before sampling them. The added physics term therefore steers the search toward operating points that are both promising and physically consistent, while avoiding unnecessary exploration of the hazardous region. For comparison, we also tested a distance-based trust-region strategy (green trajectories), conceptually similar to SafeOpt in that it restricts new evaluations to the neighborhood of previously validated points \citep{suiSafeExplorationOptimization2015}. Although this mechanism avoids unsafe exploration, it expands too cautiously and may converge slowly or remain trapped in local optima, as illustrated in Fig.~\ref{fig:traj_c}. The proposed method achieves a better compromise between safety and global exploration because the residual penalty provides additional structure beyond purely distance-based restrictions.

\begin{figure}[h]
  \centering
  \includegraphics[width=0.45\textwidth]{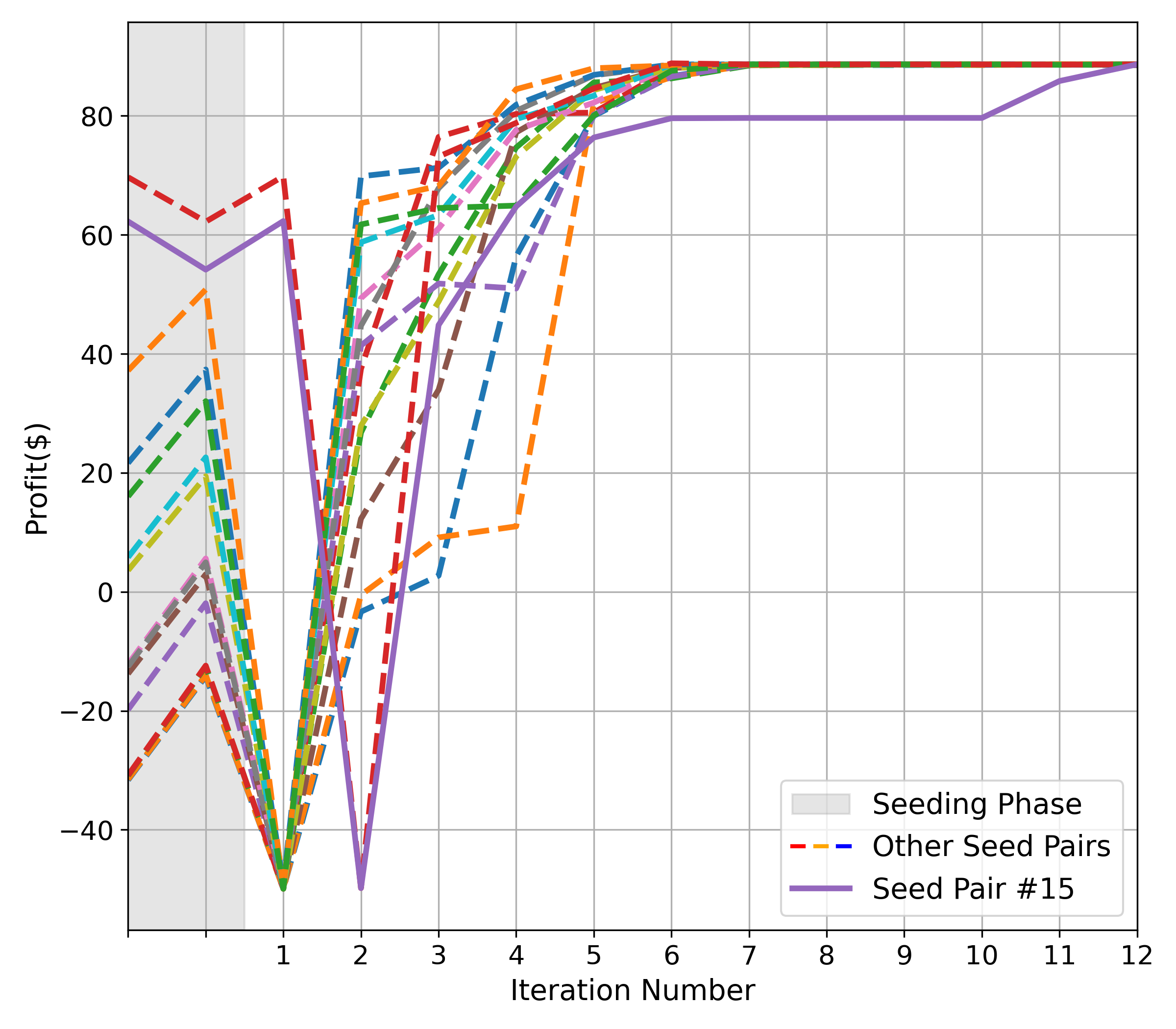}
  \caption{Convergence profiles of the economic profit ($\mathrm{J}$) over 10 iterations for 15 independent runs. The algorithm demonstrates robust convergence to global optimum regardless of initialization. Note the trajectory of Seed 15 (highlighted), which successfully escapes a local optimum to reach global maximum.}
  \label{fig:Profit_Profile}
\end{figure}

\begin{figure}[h]
  \centering
  \includegraphics[width=0.45\textwidth]{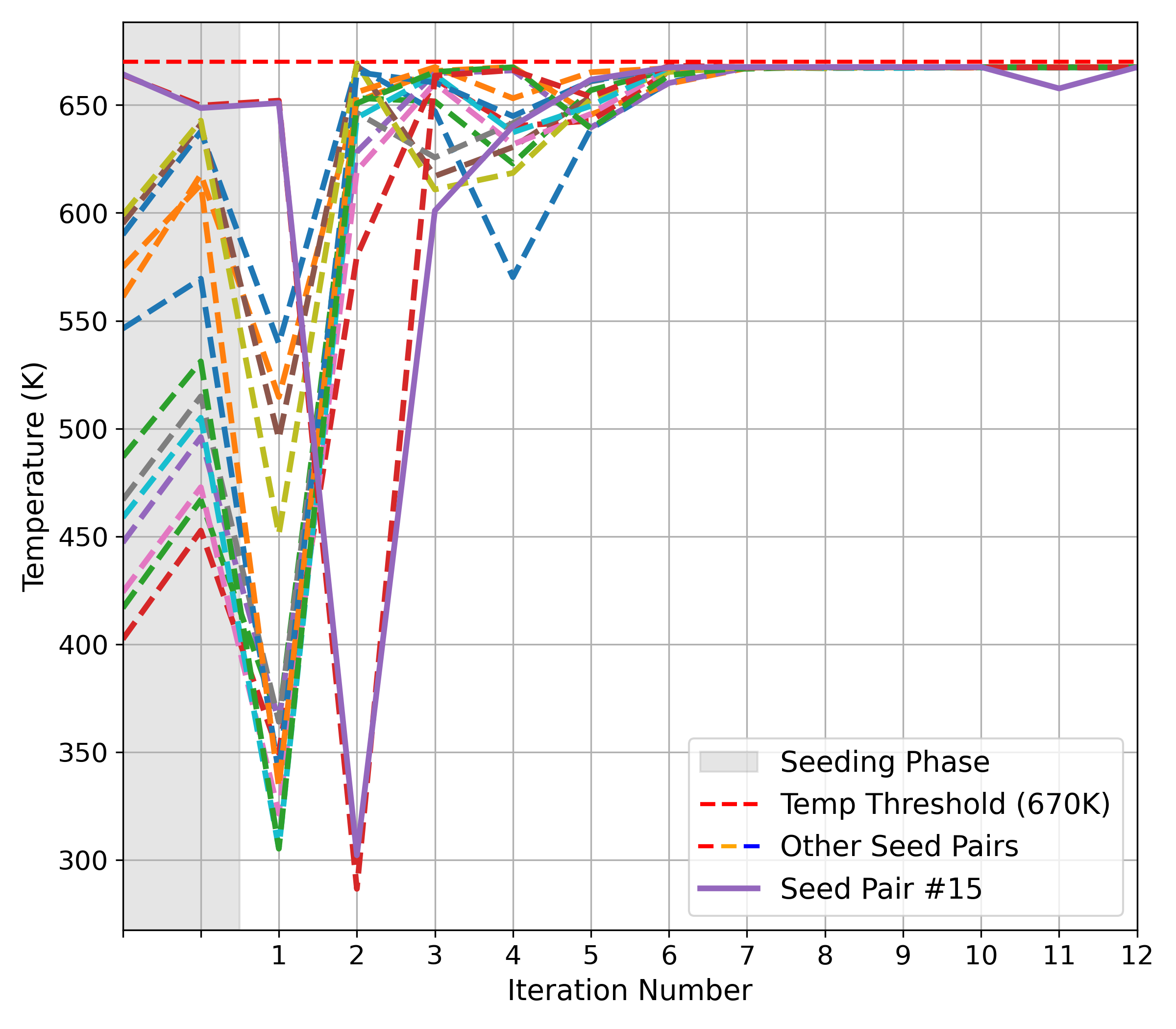}
  \caption{Evolution of the reactor temperature ($T$) for all optimization runs. The horizontal dashed line marks the critical safety constraint ($T_{\max} = 670$ K). All runs strictly adhere to the safety threshold.}
  \label{fig:Temperature_Profile}
\end{figure}

To assess robustness, we performed 15 independent optimization runs, each initialized from a different pair of safe seed points. These seed points were generated by randomly sampling the input domain and rejecting candidates whose simulated steady state violated the temperature constraint. This mimics the use of historically available safe operating data. The corresponding convergence and temperature profiles are shown in Figs.~\ref{fig:Profit_Profile} and \ref{fig:Temperature_Profile}.

Figure~\ref{fig:Profit_Profile} shows that all runs move toward the high-profit region despite different initial conditions. Most runs converge rapidly, while Seed 15 converges more slowly because it is initialized near a suboptimal local peak. Even in this more difficult case, the exploration term eventually drives the optimizer into unexplored regions and enables escape toward the global optimum.

Figure~\ref{fig:Temperature_Profile} shows that the reactor temperature remains below the safety limit in all 15 runs. This indicates that the combination of the uncertainty-aware temperature penalty and the physics-residual penalty is effective in preventing unsafe evaluations throughout the optimization.

\section{Conclusion}

This work proposed a physics-informed BO framework for safe and data-efficient RTO when a complete plant model is unavailable but selected physical relations remain known. Rather than learning the economic objective directly, the method models the steady-state reactor outputs with independent GPs and evaluates profit analytically from the GP mean predictions. Safety is handled conservatively through an upper confidence bound on reactor temperature, while physical consistency is encouraged through a penalty based on the steady-state energy-balance residual. The CSTR case study showed that the proposed method converges reliably to the economically favorable operating region from multiple safe initializations while respecting the temperature constraint throughout the optimization. Relative to a baseline BO variant without the physics penalty, the method avoids unnecessary sampling in the unsafe high-temperature region. Relative to a purely distance-based safe exploration strategy, it provides less conservative and more effective global search. These results suggest that embedding partial first-principles knowledge into the acquisition function can substantially improve BO for safety-critical process optimization problems. The same idea can be extended to other applications in which only incomplete but trustworthy physical relations are available.

\section*{Acknowledgements}
Marta Zag\'orowska acknowledges support for this research by the Delft Technology Fellowship.


\bibliographystyle{plainnat}
\bibliography{references}             
\end{document}